# Silicon Nanophotonic Waveguides for the Mid-Infrared

Tom Baehr-Jones, Alexander Spott, Rob Ilic, Andrew Spott Boyan Penkov, William Asher and Michael Hochberg

It has recently been shown that silicon nanophotonic waveguides can be used to construct all of the components of a photonic data transmission system on a single chip[i,ii]. These components can be integrated together with CMOS electronics to create complex electronic-photonic integrated circuits[iii]. It has also emerged that the high field confinement of silicon nanoscale guides enables exciting new applications, from chip-scale nonlinear optics[iv] to biosensors[v] and light-force activated devices[vi]. To date, most of the experiments in silicon waveguides have been at wavelengths in the near-infrared, ranging from 1.1-2 µm. Here we show that single-mode silicon nano-waveguides can be used at mid-infrared wavelengths, in particular at 4.5 µm, or 2222.2 cm$^{-1}$. This idea has appeared in theoretical literature, but experimental realization has been elusive[vii]. This result represents the first practical integrated waveguide system for the mid-infrared in silicon, and enables a range of new applications.

When building telecommunications systems, it is vital to operate at near-infrared wavelengths in order to maintain compatibility with existing systems. Many other types of optical systems operate in the near-infrared as a matter of convenience: A wide variety of commercial optical components are readily available at 1310, 1480 and 1550 nm, largely as a spinoff of the extensive commercial work in telecommunications. But these wavelengths are not suitable for many applications, where it is necessary to manipulate light with wavelengths in the 2-20 µm (mid-infrared) regime. A few prominent examples consist of thermal imaging (2.5 µm to 15 µm wavelengths)[viii], chemical bond spectroscopy[ix] (which often spans from the visible to 20 µm and beyond), astronomy[x], gas sensing[xi], and military applications such as missile countermeasures[xii].

Historically, the mid-infrared has been a problematic region for photonics. Coherent sources have been bulky and expensive[xiii], or have required cryogenic cooling[xiv], as did common mid-infrared detectors[xv]. Of course, the lack of integrated optical waveguides meant that mid-infrared systems were also implemented using free-space optics. Recently, the landscape has begun to change dramatically. Inexpensive, reliable, single-mode quantum cascade lasers are now available commercially all the way to 9 µm wavelengths, with powers in from 10-100 mW and near room-temperature operation[xvi]. Single-mode fibers are now available at wavelengths out to 6 µm[xvii,xviii], as are mid-infrared photodetectors with bandwidths over 1 GHz[xix]. As a result, building a single-mode optical system in the mid-infrared is now within the financial and technical reach of any modestly well-funded research group or small company. In fact, it is possible to build a full test system with decent performance for well under $100,000, which is comparable to the cost of many swept-wavelength near-infrared systems. This combination of new capabilities strongly suggests that mid-infrared photonics is poised to take off as a field in the immediate future.

But there are many missing pieces. High bandwidth modulators do not exist for these wavelengths. Neither are there low-loss splitters, tunable filters, or any of the rest of the building blocks of complex fiber-coupled photonic systems. Silicon waveguides provide an ideal platform for us to build all of these components. Furthermore, the silicon photonic platform offers the opportunity to integrate all of these components along with control electronics on the same substrate – opening the possibility of building, for instance, FTIR systems-on-a-chip with multi-centimeter path lengths but nanoliter sample volume.

The first step on this path is to construct low-loss silicon waveguides. Here we present the first measurements of functional, low-loss silicon nanoscale guides for the mid-infrared. These waveguides are built using the silicon-on-sapphire (SOS)[xx] materials system, which has been used in the electronics industry as an alternative to silicon-on-insulator (SOI). SOS is particularly desirable for this application because of the lack of a high index substrate, which eliminates the issue of substrate leakage. Mid-infrared guides could also be built using free-standing silicon guides[xxi], germanium-on-silicon, or highly customized SOI with a thick bottom oxide. SOS, in particular, has the advantage of offering the ability to build high-confinement, fully etched waveguides from 1.1 µm all the way to around 6.2 µm[xxii] – over two octaves of bandwidth, including the telecommunications region, while maintaining electronics compatibility.

The SOS waveguides were simulated using a Yee-grid based eigensolver[xxiii]. As shown in figure 1, it was found that a 1.8x0.6 µm ridge waveguide offered a small mode (around 1.1 µm$^2$), tight bend radius (10 µm with negligible losses), and allowed the silicon etch to stop on the sapphire, simplifying fabrication. The mode size is around 1/19 of a square wavelength in free space at 4.5 µm. At 4.5 µm, the

TM0 mode was expected to either not guide or be weakly confined, since it is near cutoff. As we will describe later, experimental evidence suggests that only the TE0 mode guides with low loss. The devices are terminated in waveguides with widths of 8 µm, which were coupled into the 1.8 µm waveguides with a taper. This configuration was used to increase coupling efficiency. Predicted coupling losses from the 8 µm wide guide to the single-mode mid-infrared fibers that we used were on the order of 11 dB.

The waveguides were fabricated using standard semiconductor processing techniques. Epitaxial silicon-on-sapphire wafers (100 mm diameter) were used as starting material. Photopatterning was accomplished using a stepper and patterns were transferred using a $CF_4$ plasma in an Oxford Instruments parallel-plate RIE. Resist was removed using wet resist remover, although some resist residue appeared to remain on the surface of the waveguides due to incomplete stripping. The wafer was manually cleaved through the 8 µm wide waveguide segment, leaving an optical quality edge.

The devices were tested using an Ekspla PL2241 Nd:YAG laser that drove an Ekspla PG501/DFG optical parameteric generator/difference frequency generator (OPG/DFG). The OPG/DFG produceslinearly polarized light from 2-9 µm, with 4.5 µm radiation used in our testing. This laser was coupled through a polarizer into ZnSe lens with 12.7 mm focal length, and into a 9/125 µm single-mode optical fiber S009S17 from IR Photonics. The fiber, chip and detector were held on stages with piezoelectric actuated micrometers (Newport PZA12). The output of the chip was coupled directly to free space and then into a PVI-5 detector from Vigo Systems.

Because the OPG/DFG provided 30 ps pulses of IR light at a repetition rate of 50 Hz, we used a boxcar amplifier (Stanford Research SR250) to reject signal during the times when the laser was not providing output. The signal to noise ratio was further enhanced by mechanically chopping the laser at 2 Hz, and using a Signal Recovery 7265 lock-in amplifier to detect the 2 Hz modulated signal. This resulted in an overall signal to noise ratio of 85 dB. Coupling from the free-space mode into the single-mode fiber was achieved with around 12 dB of insertion loss, leaving adequate dynamic range to perform measurements on the SOS waveguides.

In the process of testing the chip, we mechanically swept a fiber across one of the faces of the chip. We found a transmission pattern that closely matched the lithographically defined waveguide pattern, as shown in figure 2. By measuring a series of devices with different waveguide lengths on the same chip, it was possible to determine the overall waveguide loss, and to separate this from the loss of the 8 µm waveguide regions and the insertion loss in getting the light onto and off of the chip.  Devices that were tested ranged in total length from 1 mm to 1.4 cm. The waveguide losses on this chip were found to be 10.4 ± 1.2 dB/cm. A bend radius of 40 µm was used for the devices, some of which involved multiple 180 degree bends. We also included two control structures with waveguide widths of 1.2 µm, identical to other devices in all other respects. As predicted theoretically, these devices did not show any transmission at all, providing further evidence of waveguiding.

It was not possible to perfectly control the input polarization to the waveguides, due to the fact that non-polarization maintaining fiber was used. However, the polarization appeared to be fairly stable for moderate periods of time. In the case of the waveguide loss data shown in figure 3, one polarizer before the lens was held at a constant position corresponding to the peak device response, with nearly horizontally polarized light exiting the fiber. To explore the polarization of the waveguide mode, we placed a second polarizer at the output of the waveguide, and observed transmission as a function of rotation of the second polarizer, with typical results shown in figure 4. The output polarization state appeared to be independent of the input polarizer's rotation, and was also predominantly polarized in the horizontal direction. Based on this, we conclude that the waveguide likely only supports the TE0 mode.

We expect that it will be possible to significantly improve these loss numbers with improved processing. As the mode is predominantly localized in the silicon ridge, the ultimate waveguide loss should be determined by the bulk crystalline silicon losses, which are less than 0.1 dB/cm at 4.5 µm. We attribute the presently observed losses to surface roughness, and the presence of resist residue on the top of the silicon ridge.

With low-loss waveguides, it will be possible to build integrated mid-infrared lasers and detectors using techniques such as wafer bonding[xxiv] and selective-area growth, and to construct a wide variety of further devices within the silicon platform. It should also prove possible to build high-confinement integrated nonlinear optical devices, such as integrated OPO's and difference frequency generators. Another interesting opportunity emerges from the limits of lithography: A 20 nm trench represents a small fraction of a wavelength at 1.55 µm, but represents a significantly smaller fraction for light at 4.5 µm; mid-infrared waveguides may turn out to be an ideal 'playground' for exploring ideas in ultra-subwavelength photonics. In the long run, we anticipate fully integrated mid-infrared optical systems, which will be much smaller and cheaper that contemporary systems.

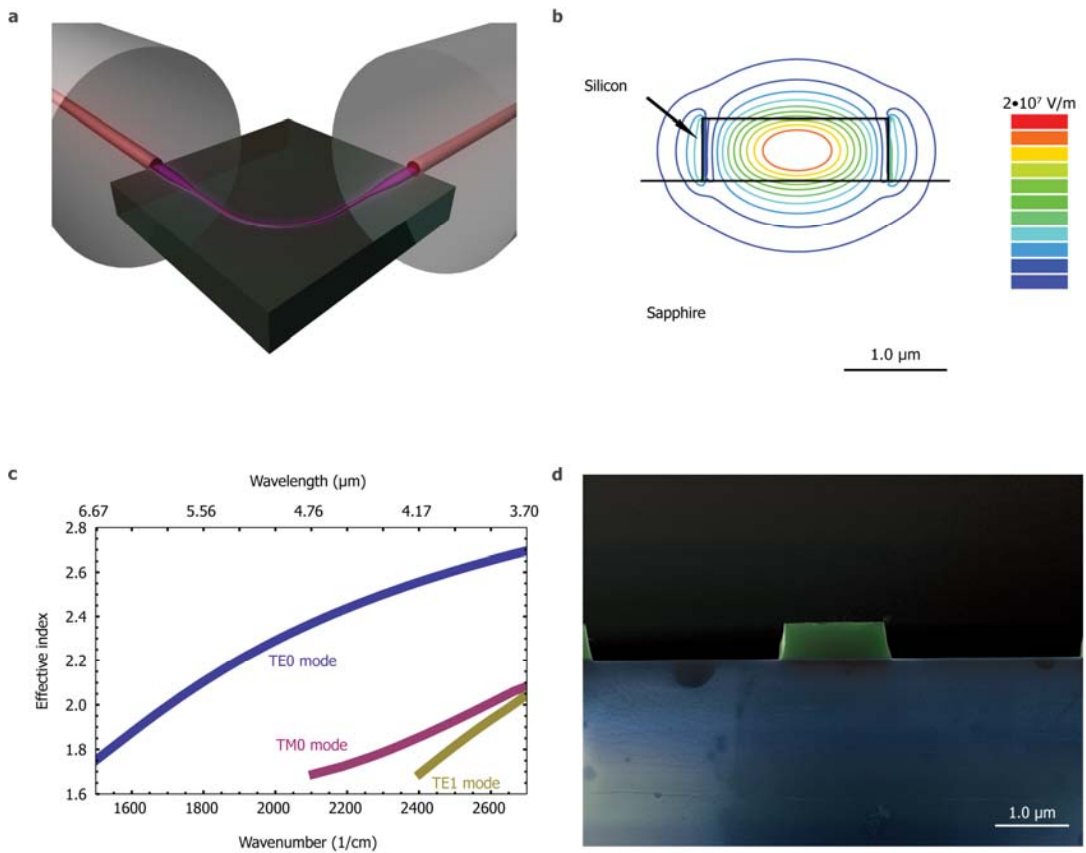

**Figure 1 Waveguide Design. a**, Three-dimensional rendering of one of the mid-infrared waveguide devices, showing the optical mode propagating through the waveguide. The rendering includes single-mode fibers butt-coupled to the input and output facets. **b**, A contour plot of the optical mode of the waveguide is shown, with the electric field magnitude corresponding to 1 Watt of average power. **c**, The dispersion diagram for the silicon-on-sapphire waveguides, calculated using a Yee grid based eigensolver. **d**, A false-color scanning electron micrograph of the cleaved endfacet of a waveguide. Silicon is shown in green, and sapphire in blue.

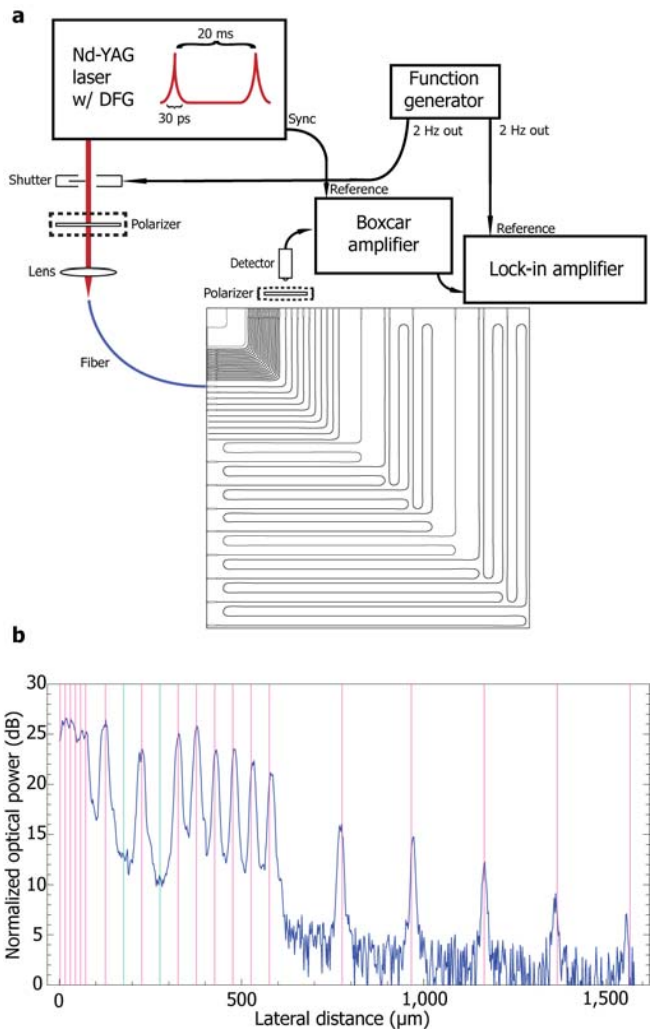

**Figure 2 Experimental Layout.** **a**, A schematic of the experimental setup used to measure the waveguides is shown. Also shown is an image of the lithographic pattern defined on the chip. Some structures consist of a single bend, while some have several bends. **b**, The optical transmission is plotted as a function of lateral fiber position, with purple lines corresponding to the positions of waveguides. These positions are identified by comparing the stage micrometer offset to the lithographically defined position, and were confirmed visually. Two vertical green lines show control structures, which had narrower waveguides and thus did not guide.

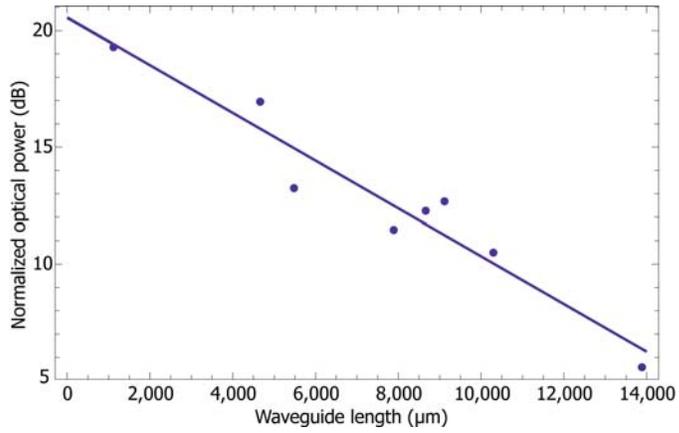

**Figure 3 Waveguide Loss Measurements**. Relative device transmitted power in dB as a function of device length for a number of bend devices, with a best fit line, showing waveguide losses of 10.4 ± 1.2 dB/cm.

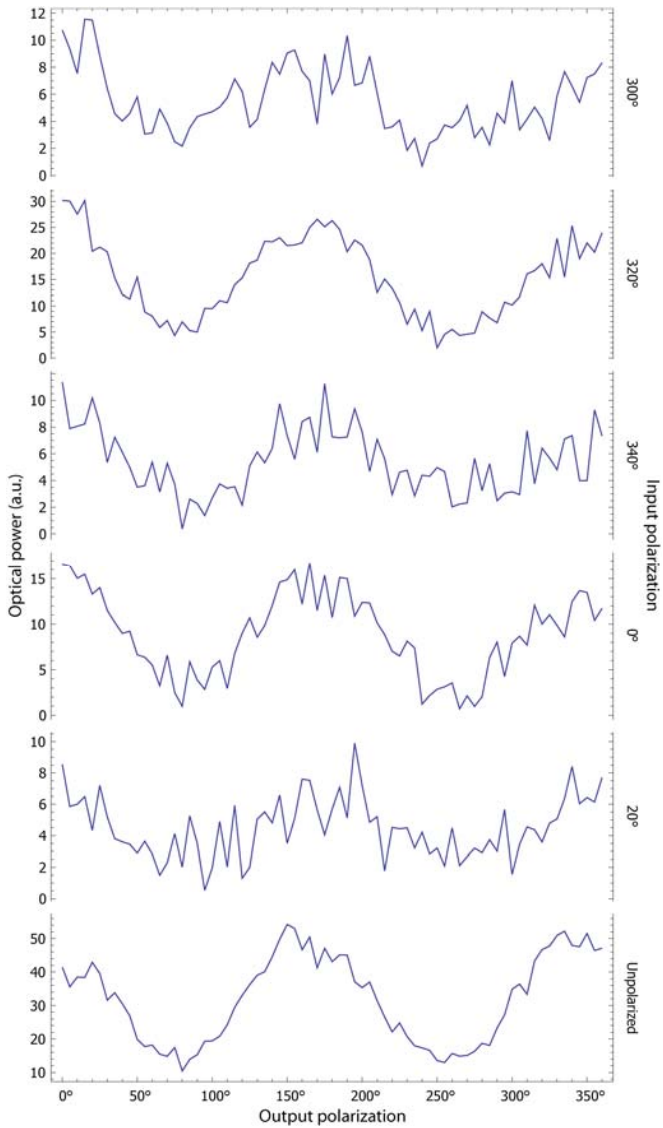

**Figure 4 Polarization Measurements.** Transmitted power as a function of the polarizer directly in front of the detector, with results shown for several positions of the initial polarizer. The absence of any shift in the location of the peaks suggests that the waveguide is functioning as a linear polarizer. To get a stronger signal, this data was taken for a group of closely spaced devices originally intended as fiducials; light is likely propagating through two or three waveguides for these results. Similar results were seen for a single waveguide device.

Acknowledgments:

The authors would like to acknowledge use of the facilities at the Cornell Nanoscale Facility for the device fabrication portion of this work. In addition, they would like to thank Gernot Pomrenke for his support through the AFOSR YIP program, the NSF STC MDITR Center at the University of Washington, Tektronix Corporation, and the Murdock Foundation. M. Hochberg would like to thank Marko Loncar for valuable discussions.